\documentclass{PoS}

\usepackage{epsfig}
\usepackage{multicol}
\usepackage{wrapfig}
\usepackage{pstricks,pst-grad}

\newcommand{\beq}{\begin{equation}}
\newcommand{\eeq}{\end{equation}}
\newcommand{\bea}{\begin{eqnarray}}
\newcommand{\eea}{\end{eqnarray}}

\newcommand{\nn}{\nonumber}
\newcommand{\e}{\hbox{e}\,}
\newcommand{\tr}{\hbox{tr}}
\newcommand{\Tr}{\hbox{Tr}}
\newcommand{\I}{\hbox{Im}}
\newcommand{\R}{\hbox{Re}}

\title{Complex Langevin simulation for QCD-like models}

\ShortTitle{Complex Langevin simulation for QCD-like models}

\author{Gert Aarts$^{1}$, Lorenzo Bongiovanni$^{1}$, 
        Erhard Seiler$^{2}$, D\'enes Sexty$^{3}$, 
\speaker{Ion-Olimpiu Stamatescu}$^{3,4}$ %
         \thanks{Partial support by the Deutsche Forschungsgemeinschaft 
	 is thankfully acknowledged}\\
	 $^{1}$Department of Physics, College of Science, Swansea 
	 University, Swansea, United Kingdom \\
	 $^{2}$Max-Planck-Institute for Physics (Werner-Heisenberg-Institute),
	  M\"unchen, Germany \\
        $^{3}$ITP, University of Heidelberg, Germany\\
	$^{4}$FEST, Heidelberg, Germany \\
        E-mail: \email{stamates@thphys.uni-heidelberg.de}}

\abstract{We first test the Complex Langevin method (CLE) on various 
simple models. We then introduce the method
of Gauge Cooling to control the dynamics of the process 
and ensure thin distributions in the imaginary direction. 
We finally apply CLE with gauge cooling to a QCD-related 
lattice model (HQCD) and compare the results by CLE and by a 
refined Reweighting method (rRW). Very good agreement is found 
in all regions of physical interest. }

\FullConference{31st International Symposium on Lattice Field Theory - LATTICE 2013\\
		July 29 - August 3, 2013\\
		Mainz, Germany}

\begin{document}

\section{Motivation and program}

The Complex Langevin Equation (CLE) has the potential
to simulate lattice models for which  usual importance sampling 
 fails. In many cases, especially for QCD at 
non-zero density, the
CLE in principle provides the (only) 
model independent procedure.

The real Langevin Equation (LE) is a well studied stochastic process.
Its redefinition as CLE is more involved. To  develop it to a reliable method
 is  both rewarding and tough. Our program
is:\\
 - Define and study the properties of the CLE, test CLE for simple models.\\
 - Apply CLE to realistic models aiming at 
 full QCD at non-zero chemical potential
  \cite{den13}.

\section{ The Langevin equation for real models}

The LE  for a real field 
$\varphi(x)$ evolving in the process time $t$ ("Langevin" time, 
here discretized) is:
{\bea
&&\delta \varphi(x;t) 
 = K[\varphi(x;t)]\,
\delta t +
\eta(x;t) \nn\\
&&\langle \eta(x,t)\rangle = 0,\ \ \langle \eta(x,t)\eta(x_1,t_1)
\rangle =
2\,\delta t \, \delta_{x,x_1}\, \delta_{t,t_1} \nn
\eea}
($\delta t$: time step, Ito calculus) with the associated Fokker-Planck equation (FPE)
\bea{\partial_t}P(\varphi,t)= \partial_\varphi \left(\partial_\varphi 
- K \right)\,P(\varphi,t) .
\eea
If the drift $K = - {\partial_{\varphi}}\, S$ with  $S$ a positive definite action
we then have asymptotically
\bea
t \rightarrow \infty \quad \quad  P(\varphi,t) \rightarrow P_{as}(\varphi) = \frac{1}{Z}\exp{(-S)}\, , \quad  Z = \int[d \varphi]\,\exp{(-S)}\, .
\eea
For positive measure the LE is
well defined and comparable with Monte Carlo. 
In the presence of a sign problem LE may have difficulties. 
One can study
this  in simple models and devise systematic cures \cite{en13} 
 overcoming old "disasters" \cite{amb}. The problems, however, 
may be inherited in the CLE.

\section{Set up for the CLE}

For a complex action the drift is also complex and this
automatically provides an imaginary part for the field.
This implies setting up the problem in the complexification 
of the original manifold  $R^n \longrightarrow
C^n$
   or $SU(n) \longrightarrow SL(n,C)$ .
The CLE then amounts to  two related, real LE with independent 
noise terms - 
here for just one variable $x \rightarrow z = x + i\, y$ and with
$K= - \partial_z S(z)$:
 \bea
{\delta z(t)} &=& K(z)\,\delta t + \sqrt{N_R}\,
\eta_R+ {\rm i}\,\sqrt{N_I}\,\eta_I \nn \\
{\rm i.e.} &&{\delta x(t)} = {\R}\,K(z)\,\delta t + 
\sqrt{N_R}\,\eta_R (t) \, , \quad{\delta y(t)} = 
{\I} \,K(z)\,\delta t +
\sqrt{N_I}\,\eta_I (t) \nn \\
  \langle \eta_R\rangle &=& \langle\eta_I\rangle =0\,,\  
  \langle \eta_R \eta_I\rangle =0 
 \, , \quad\langle \eta_R^2\rangle =\langle \eta_I^2\rangle = 
 2\,\delta t\,,\ 
\, \ N_R - N_I =1\nn
\eea
The probability distribution $P(x,y;t) $ realized in the process 
evolves according to a real FPE:
\bea
\partial_t  P(x,y,t) =L^T P(x,y,t)\, ,  \quad L =
(N_R\partial_x + {\R} K(z))\partial_x + ( N_I\partial_x + {\I} K(z))\partial_y
\eea

One can also define a complex distribution $\rho(x,t)$  
\bea
\partial_t \rho(x,t) = L^T_0 \rho(x,t)\,, \quad  L_0 = 
(\partial_x + K(x))\partial_x   \nn
\eea
with the asymptotic solution $\rho(x)\simeq \exp(-S(x))$ and 
formally prove for  {\it analytic} observables $O(z)$
\bea \quad
\int {O}(x+iy) P(x,y;t) dxdy = 
\int  {O}(x) \rho(x;t) dx . \nn
\eea
The formal proof has, however, loopholes related among others 
to a too wide $P(x,y,t)$ in $y$ \cite{trust}. 
This width may be enhanced by 
an imaginary part in the noise, therefore one usually uses
 $N_I = 0$.

\section{ Problems and models}

Very many studies for CLE have appeared since the original 
papers of Parisi and of Klauder  \cite{parkl} including critical 
analysis, cf e.g. \cite{amb}. 
The problems studied in our group include: 
Real time simulations, 
 Chemical potential, 
 $\theta$-term. We here address QCD with chemical 
 potential:
\bea
Z &=&  \int DU\, \det W\,\e^{-S_{YM}}, \label{e.zqcd}\\
W &=& 1 - \kappa\sum_{i=1}^3 \left( \Gamma_{+i} U_{x,i}T_{i} 
+ \Gamma_{-i} U_{x,i}^{-1} T_{-i} \right) 
-\kappa \gamma \left( \e^\mu \Gamma_{+4} U_{x,4}T_{4} +
\e^{-\mu}\Gamma_{-4} U_{x,4}^{-1} T_{-4} \right)
\nn
\eea 
 Wilson fermions, $T$: lattice translations, $\Gamma_{\pm \mu} = 1 \pm 
\gamma_\mu$, $\kappa = 1/(2M+8)$, $M$ bare mass,
$\gamma$ bare anisotropy. The temperature is introduced as
$aT=\frac{\gamma}{N_\tau}$. We have $\det W(\mu) = [\det W(-\mu)]^* $.
CLE does not have an overlap problem such as the 
reweighting methods (RW) and does not involve approximations like
expansion methods: The ensemble is generated at the actual values
of the parameters without restriction in the latter.
The problems one encounters with CLE can be:

1) Accumulation of numerical errors. Typical effect: run-aways, 
divergence of some quantities. This
can be efficiently solved by adaptive step size 
(which we use throughout our analysis) \cite{adap}. 

2)
 insufficient fall off of $P(x,y)$ in the non-compact directions  can
lead to imprecise sampling and also can 
spoil the formal proof of equivalence. This 
 can be to a good extent 
approached by controlling the dynamics of the process, e.g.
by {\it gauge cooling} for gauge models (see below). 

3) 
 Non-holomorphy of the drift 
can invalidate the formal proof of equivalence, e.g.
poles of $K(z)$ coming from zeroes of the measure, $\rho(z)$.
To this challenge we  only have partial answers \cite{en13}.

To control the reliability we implement a number of 
checks (Consistency Conditions CC \cite{CC} - 
combinations of observables which should vanish identically 
in the correct case -, 
monitoring the 
 distributions) and stabilizing procedures generalizing the 
CLE \cite{stab}.

\section{One link effective model}

A paradigmatic effective model is an SU(3) model with one link $U$. 
Diagonalizing $U$ we obtain a reduced model with the reduced Haar measure
in the three diagonal exponents $w_i$
\bea
Z &=& \int [dw] \rho(w) \, , \quad \rho(w) ={\e}^{-S_{YM}}\,H(w)\, D(w)\, 
{\tilde D(w)} \, , \quad w_1+w_2+w_3 = 0 \\  
 &&S_{YM} = - \frac{\beta}{2} \sum_{i=1}^3 \left( \alpha_i e^{i w_i} +
\frac{1}{\alpha_i} e^{-i w_i} \right) \, , \quad H = \sin^2 \frac{w_2 -
 w_3}{2} \sin^2 \frac{w_3 - w_1}{2} 
\sin^2 \frac{w_1 - w_2}{2} , \nn \\
 &&D =  1 + C \tr U + C^2 \tr U^{-1} + C^3 , \ \ C  =  2 \kappa {\e}^{\mu} 
 \,; \quad
{\tilde D}  =    1 + {\tilde C} \tr U^{-1} + 
{\tilde C}^2 \tr U + {\tilde C}^3 , \ \ {\tilde C} = 2 \kappa {\e}^{-\mu} \nn\\
&&K_i(w) = K_{YM,\,i} + H^{-1}\partial_{w_i} H   +
D^{-1}\partial_{w_i}D  +  {\tilde D}^{-1}\partial_{w_i}{\tilde D} .\eea
The $\alpha$'s simulate the "staples" of the neighbours. As a general remark,
 observe that 
one first needs to complexify the variables, here in going from su(3) to sl(3,C),
before deriving the drift $K(w)$. 
$K(w)$ is generally  meromorphic due to possible
poles from 
  the zero's of the determinants. 
  
\begin{figure}[h]
\begin{center}
\includegraphics[width=0.45\columnwidth]{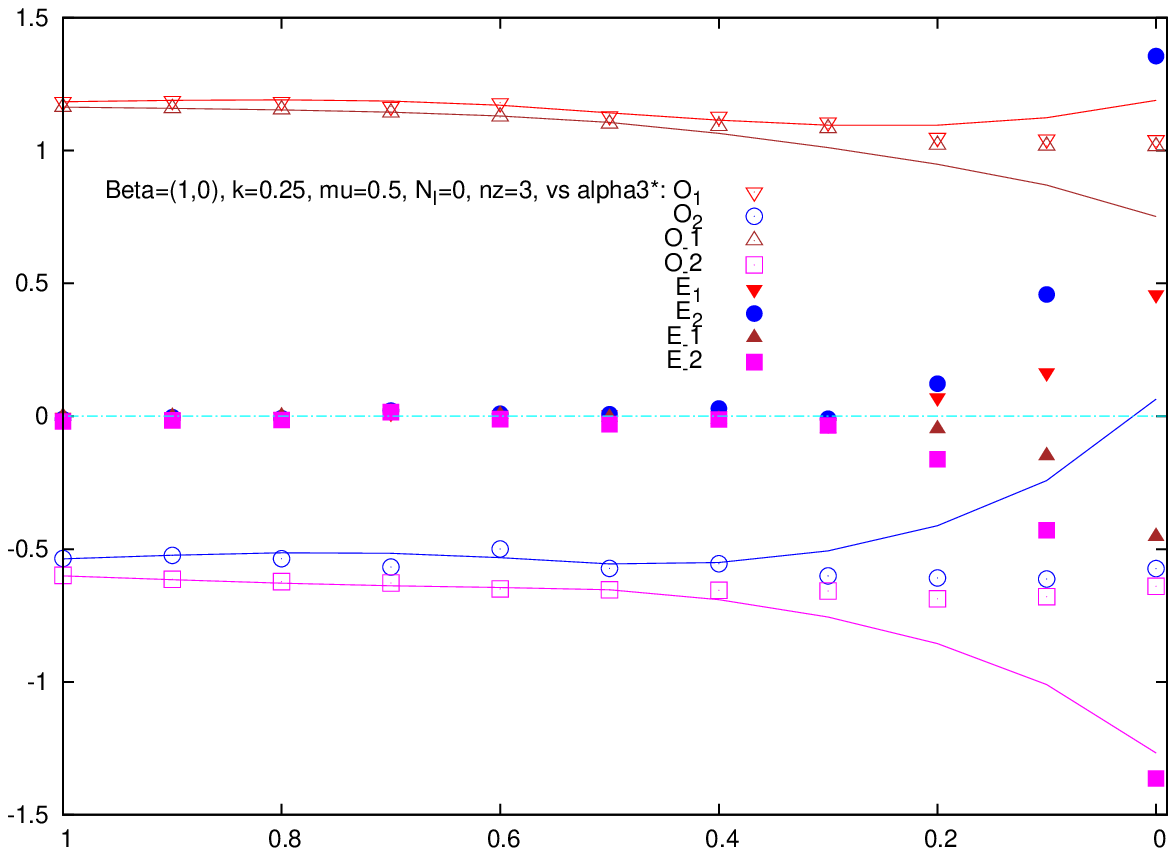}  
\includegraphics[width=0.45\columnwidth]{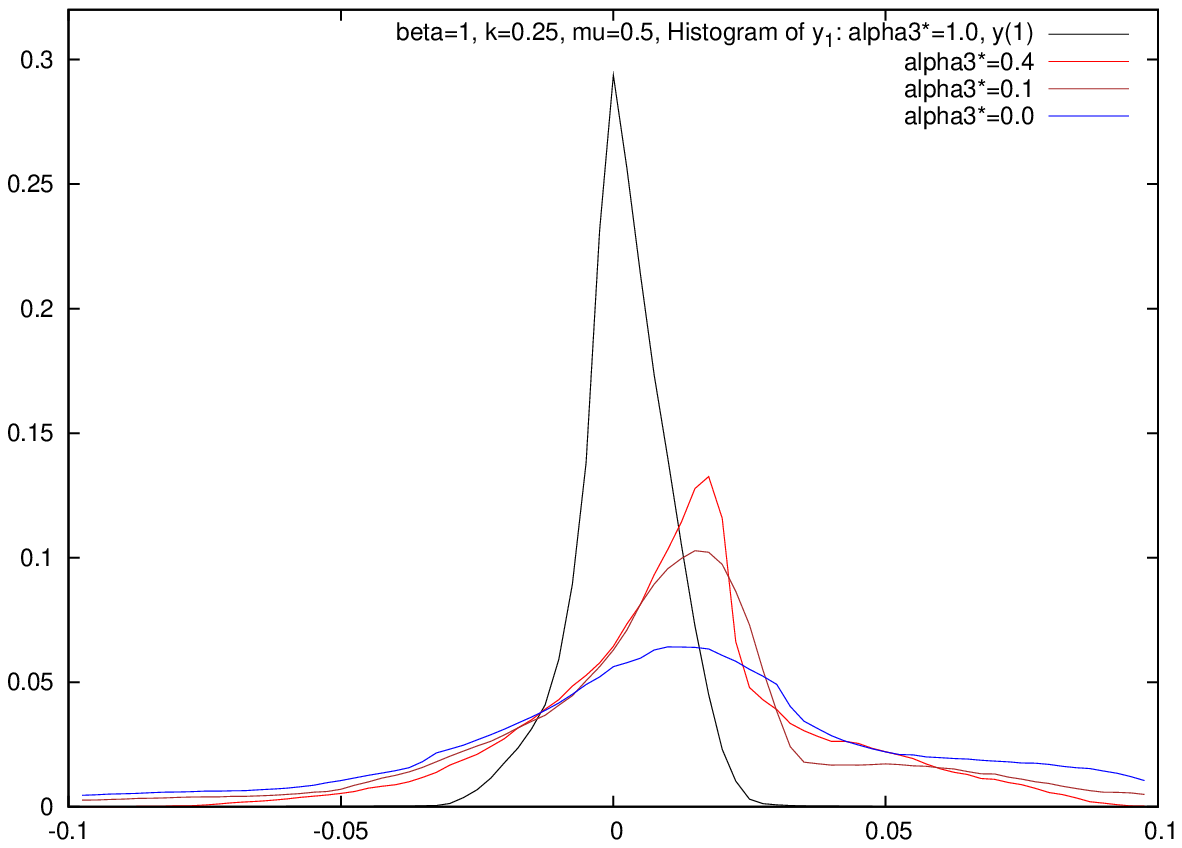}  
\caption{Effective model: Observables $O_q=e^{iqz}$, CC's $E_q$, $q=\pm1,\pm2$ vs   
 $Re \alpha_i$ and the {\it y}-distribution.  }
\label{effmod}
\end{center}
\end{figure}

  We  find that correct results are obtained if the flow does not 
  drift too 
  far in the non-compact directions - Fig. \ref{effmod}. This effect must
   be monitored and suggests possibilities to redesign the process to
    control the skirt of the distribution.
    For $\alpha$ complex, far from $1$, CLE departs from the exact 
results (left plot, solid lines). This correlates here with wide skirts of the
 $ {\I} w$-distributions (right plot). 
 
  Another source of non-reliability 
    are the poles. This is seen 
  in this model \cite{en13} as well as in more complex models 
  where the determinants may have zeroes \cite{kim}. 
\section{Many Links models and Gauge Cooling.}

To see the effect of many variables we consider an exactly soluble 
Polyakov chain model:
\bea
 -S= (\beta+2\, \kappa\e^\mu )\, P +
(\beta+2\, \kappa\e^{-\mu} )^*\, 
 P^{-1} \nn
 \eea
 with ${\rm P}= \Tr\left( U_1 \cdots U_N \right)$, $N$ up to $ 1024$.
  The process runs in all $8 N$ (complex) "angles" $A^a_i$:
 \bea \delta A^a_{i} = \epsilon  K^a_{i}(U) + 
 \sqrt{\epsilon}\,\eta \ , \ \  
 U_{i} \rightarrow \e^{ i\,\sum_a \lambda_a\, \delta A^a_{i} }
 \, U_{i} \nn
 \eea

\begin{figure}[h]
\begin{center}
\includegraphics[width=0.45\columnwidth]{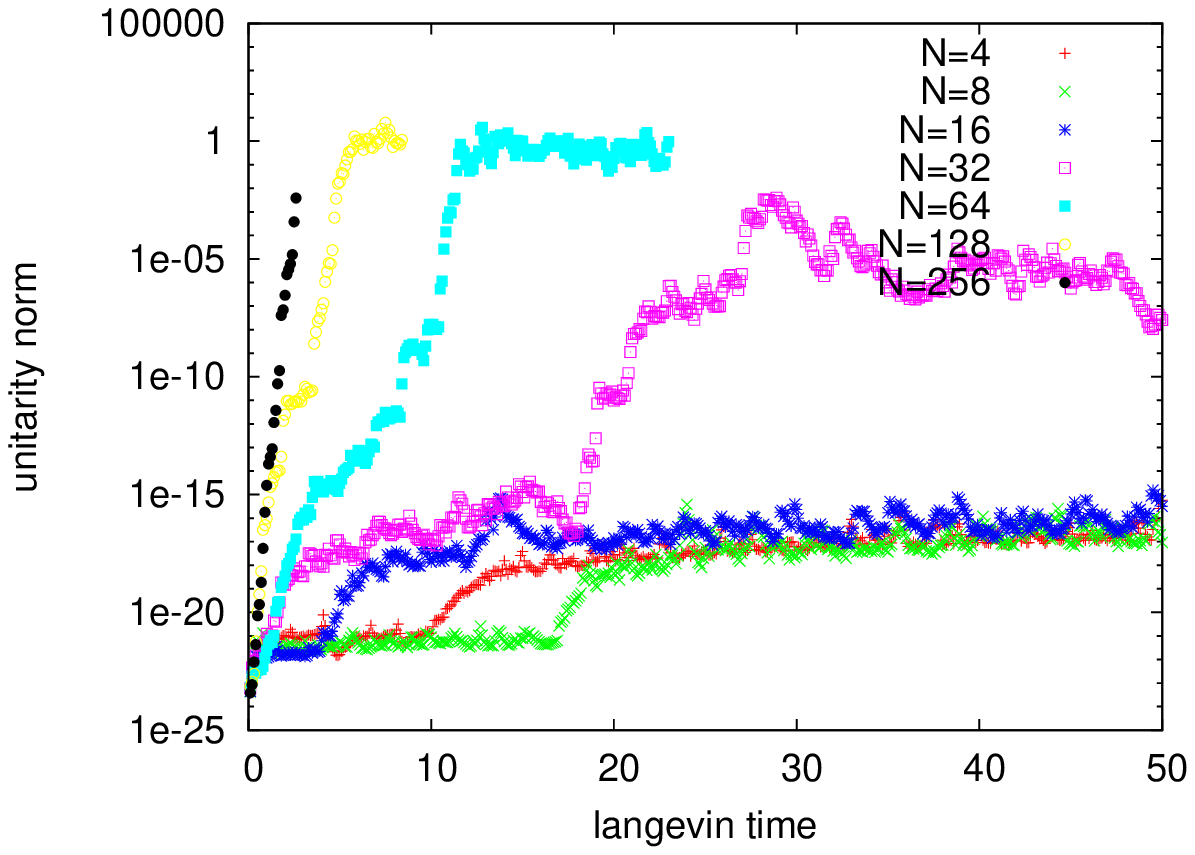}  
\includegraphics[width=0.45\columnwidth]{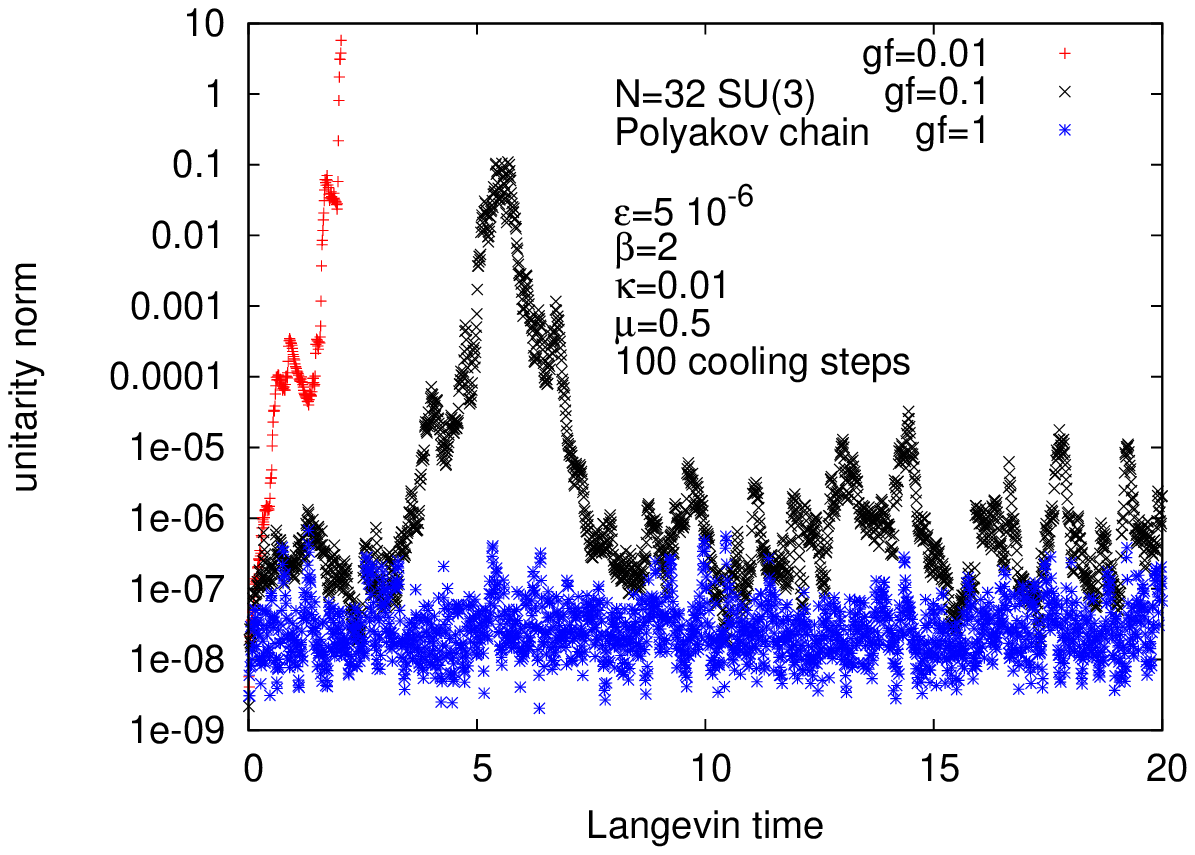}  
\caption{Polyakov chain model, evolution of the unitarity
 norm in Langevin time. Left plot:
$\mu=0$, various chain lenths, no cooling. Right plot:   
$\mu >0 $, $N=32$, various coolings.  }
\label{polch1}
\end{center}
\end{figure}
\begin{figure}[h]
\begin{center}
\includegraphics[width=0.45\columnwidth]{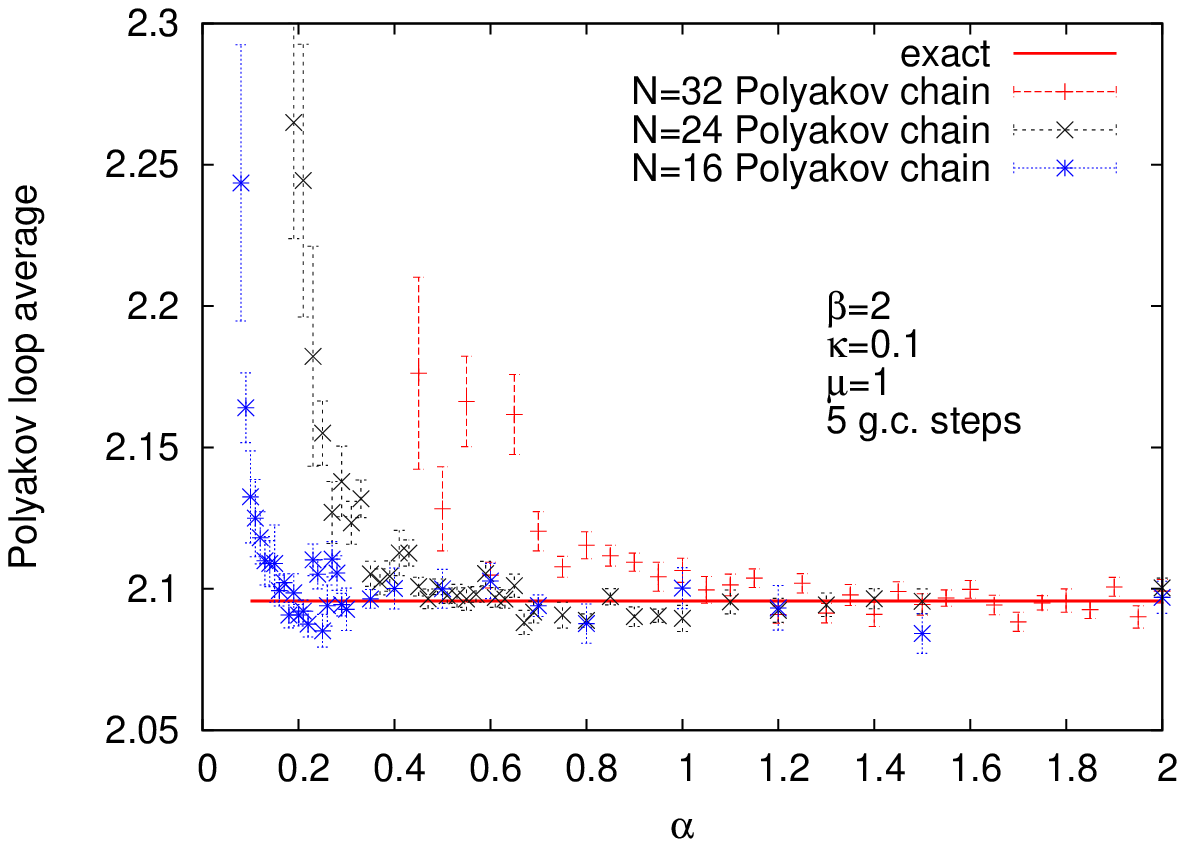}  
\includegraphics[width=0.45\columnwidth]{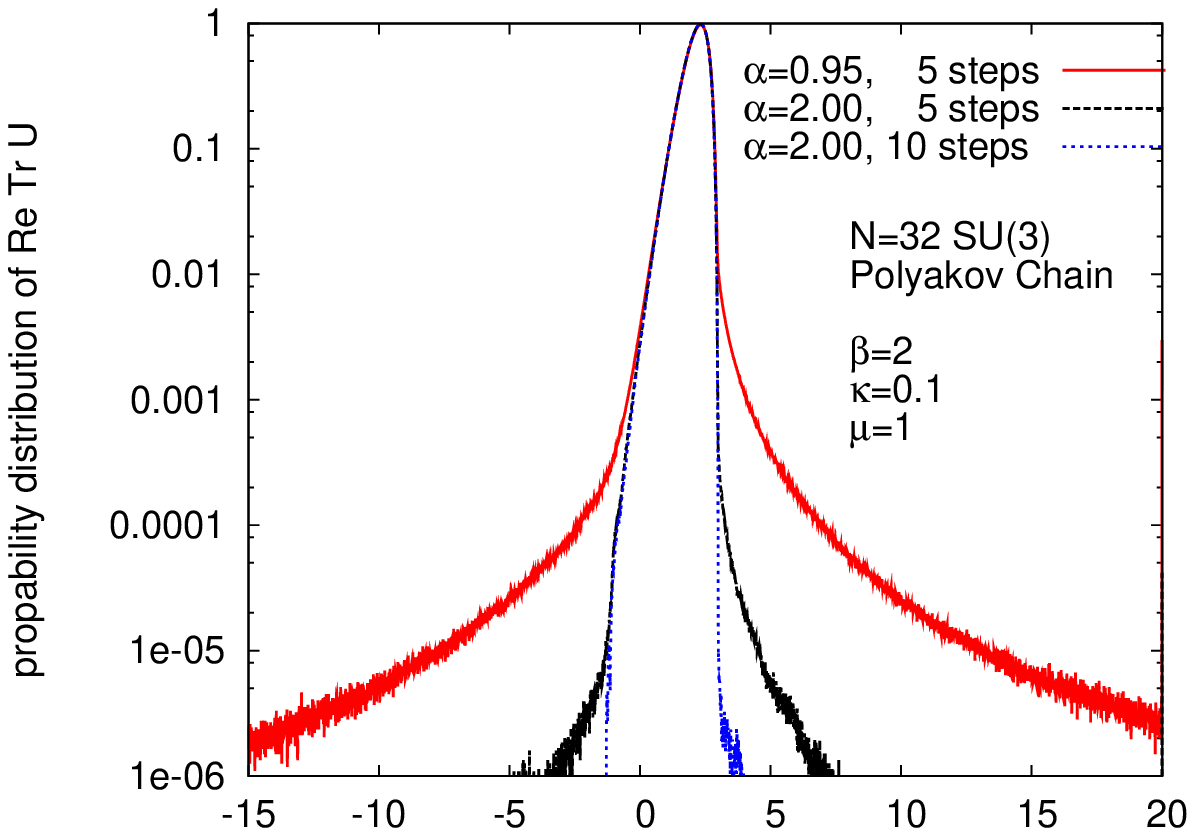}  
\caption{Polyakov chain model. Left plot: Polyakov loop average 
for various chain lengths as function of the cooling strength.
 Right plot:  Polyakov loop distribution for different amount of cooling.  }
\label{polch2}
\end{center}
\end{figure}


 with holomorphic drift $K$. For 
 large $N$ we observe, however, wrong evolution even in the real
 case ($\mu = 0$) if we set up the process as CLE, 
 although the drift and noise are real! We quantify
 this by measuring the departure of the links from unitarity with
 a {\it unitarity norm}, e.g.
 \bea {\cal U} = \sum_{links} \left[\frac{1}{2} \Tr \left(U\,
  U^{\dag}+U^{-1}\,U^{-1\, \dag} 
 \right)-3\right].\eea
  This effect - Fig. \ref{polch1}, left plot - suggests that numerical 
 imprecisons may trigger unstable modes leading away from the real axis. For simpler models 
  fixing the gauge was observed to help
  \cite{bersex}.
Using the gauge symmetry of the problem we now define a general {\it Gauge Cooling} 
procedure to bring the system as near as possible to the 
unitary manifold.
 This proceeds by successive non-compact gauge transformations
 along the gradient of the unitarity norm $\cal U$ on the gauge orbits
 \bea
  R_k = e^{-\alpha\, \epsilon\, d{\cal U}}\ , \quad U_k \rightarrow R_k\,  U_k \, , 
  \quad
  U_{k-1} \rightarrow U_{k-1} R_k^{-1} 
  \eea
with
 $\alpha$:  the strength of the {\em gauge force}, $\epsilon$: Langevin step size.
For $\mu > 0$ ${\cal U}$ should not be 0 but stabilize. This we see after 
gauge cooling (large $\alpha$ and/or many cooling steps)
 - Fig. \ref{polch1}, right plot. Then also the results are correct and the  
 non-compact
distributions narrow - Fig. \ref{polch2}.

\section{Heavy  QCD at non-zero chemical potential.}

Taking in the hopping parameter 
expansion of the fermionic determinant the double limit \cite{bender}
\bea
\kappa \rightarrow 0, \  \mu \rightarrow \infty , \ \ \zeta = \kappa \,
\e^{\mu} \, : fixed\, . \label{e.hadmlim}
\eea
we obtain an approximation for QCD at large mass and chemical potential.
In this limit only the Polyakov loops 
survive and the determinant factorizes. This can be used, e.g.
in refined reweighting (rRW) simulations (cf \cite{pietri}, where also the relevant formulae and
the next corrections are given). 
Using also the inverse Polyakov loops (which in the above limit are 
not present) one obtains a model by itself, HQCD,  which can be followed also
 away from this limit \cite{aarsta}. Both a CLE and an accordingly
 "symmetrized" 
 rRW  approach can be implemented for this model.

\begin{figure}[h]
\begin{center}
\includegraphics[width=0.45\columnwidth]{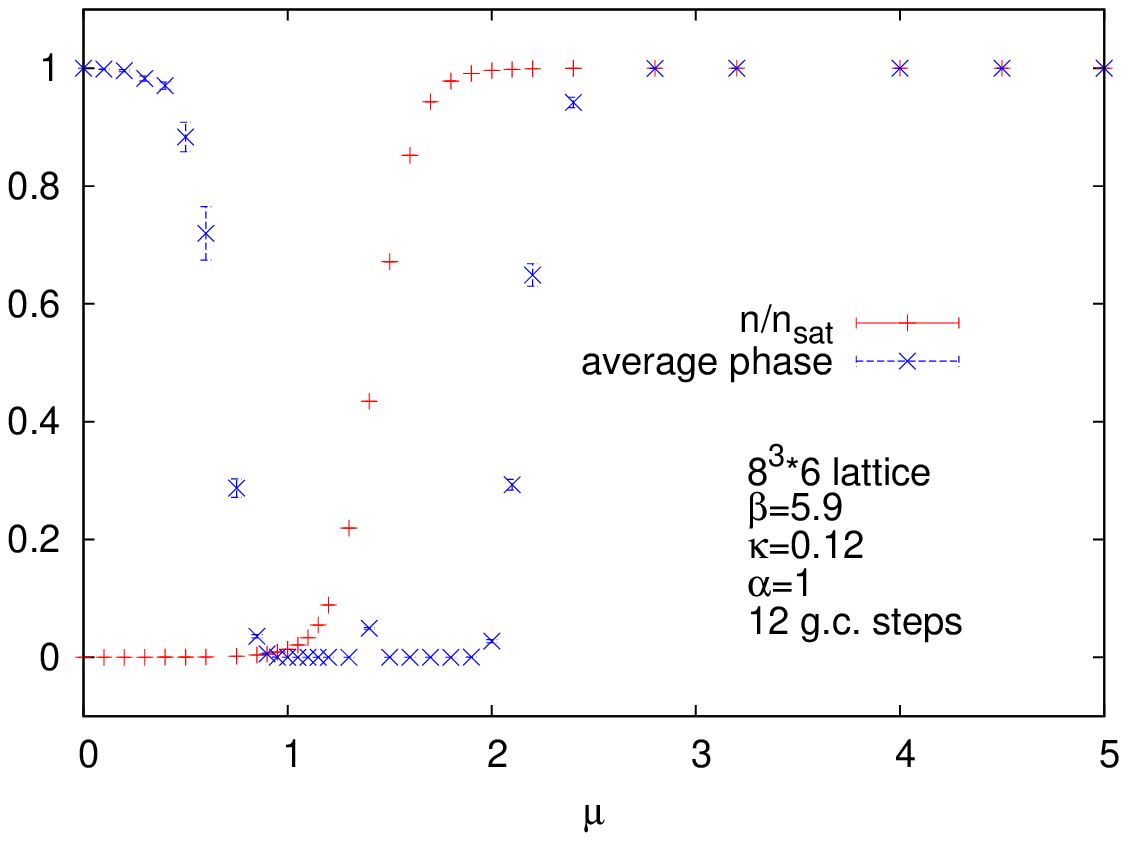}  
\includegraphics[width=0.45\columnwidth]{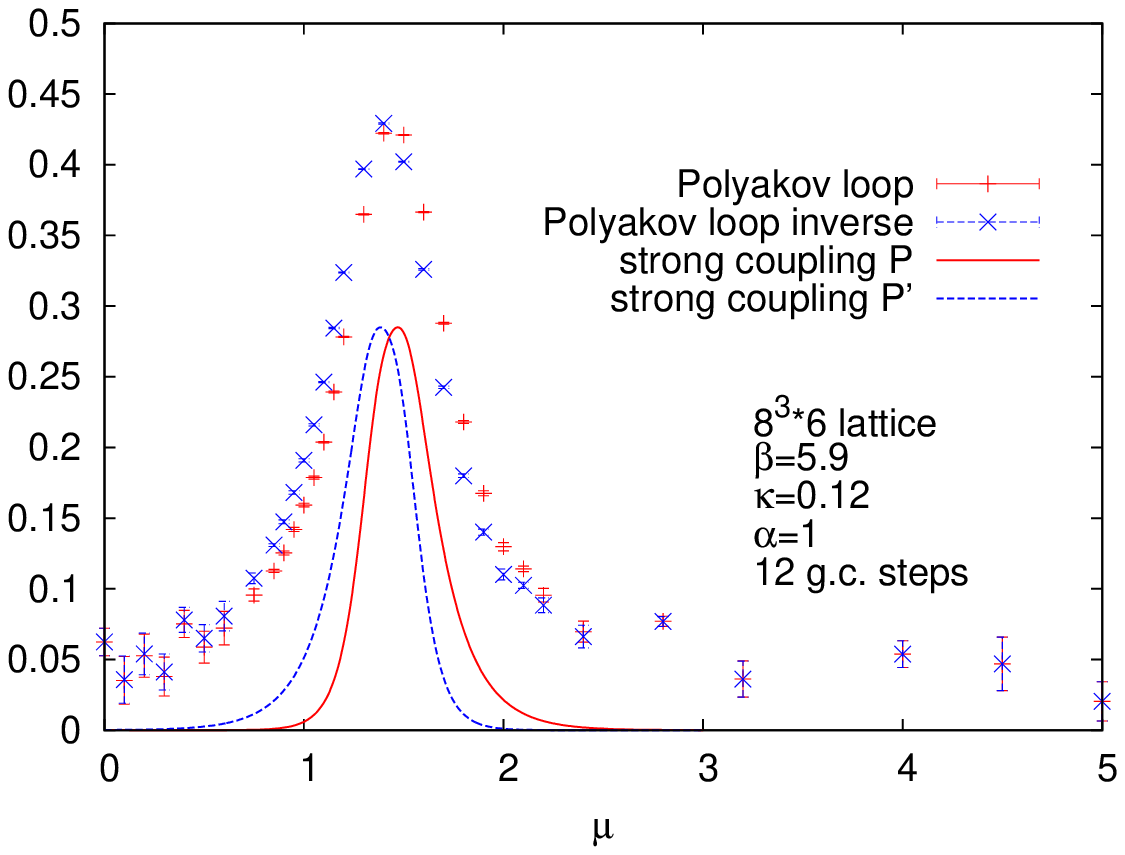}  
\caption{HQCD model, CLE with gauge cooling: Baryon density 
and average phase (left plot) and
Polyakov and inverse Polyakov loop averages (right plot) vs $\mu$, $8^3 6$ lattice.  }
\label{hqcd1}
\end{center}
\end{figure}

Using CLE we observe for HQCD the same effects as for the Polyakov chain. The 
following results are obtained with gauge cooling, which ensures a 
stabilized unitarity norm \cite{sss}.
We measure plaquettes, Polyakov loops 
$\rm P$ and ${\rm P}^{-1}$, baryon density $n$ and the
average phase:
\bea\left\langle \exp(2i\phi) \right\rangle \equiv 
\left\langle \det {\bf M}(\mu)\,
\det{\bf M}(-\mu)^{-1}\right\rangle .\eea 
The full YM action is used. 
We observe stable results for all $\mu$ all the way from $\mu = 0$ up 
to deeply in the saturation regime - Fig. \ref{hqcd1}. The results 
show the expected behaviour of the Polyakov loops and baryon 
density and that the method work very well also in the region 
where the phase factor is practically 0 (see also \cite{phip} 
for resummed strong coupling results).

\begin{figure}[h]
\begin{center}
\includegraphics[width=0.45\columnwidth]{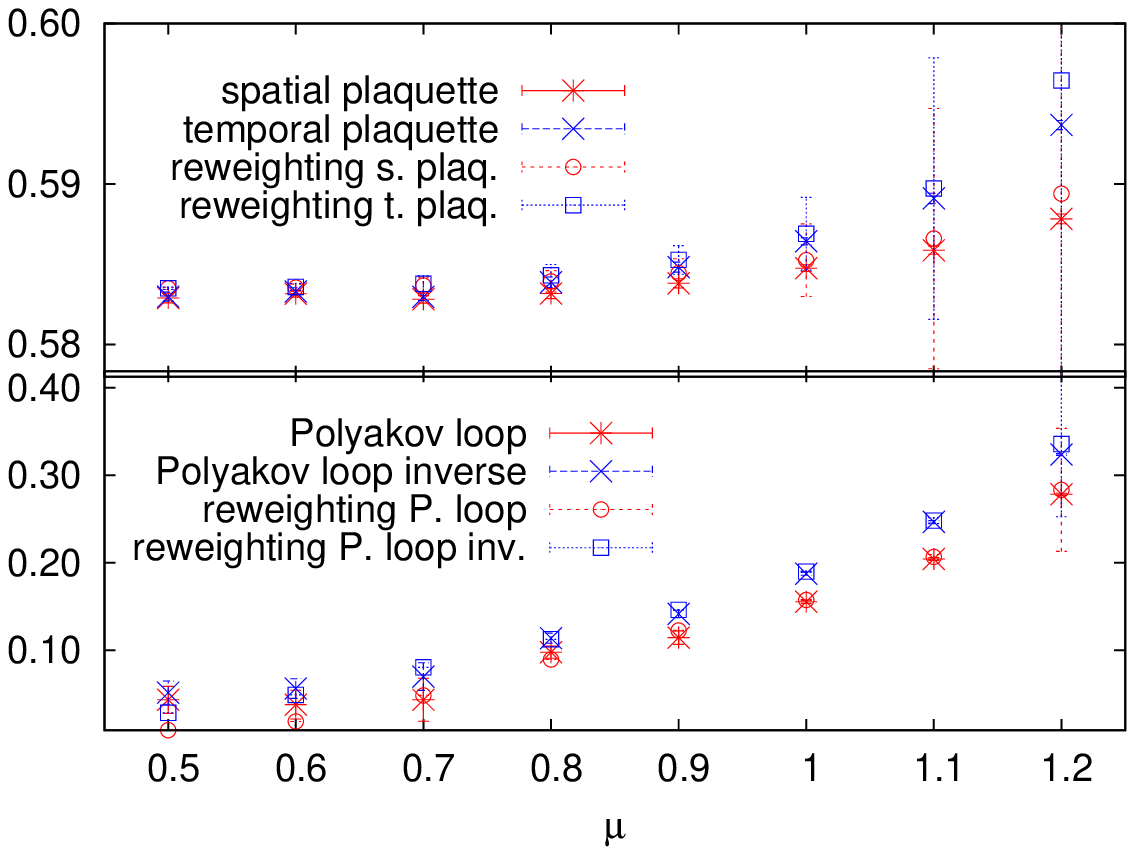}  
\includegraphics[width=0.45\columnwidth]{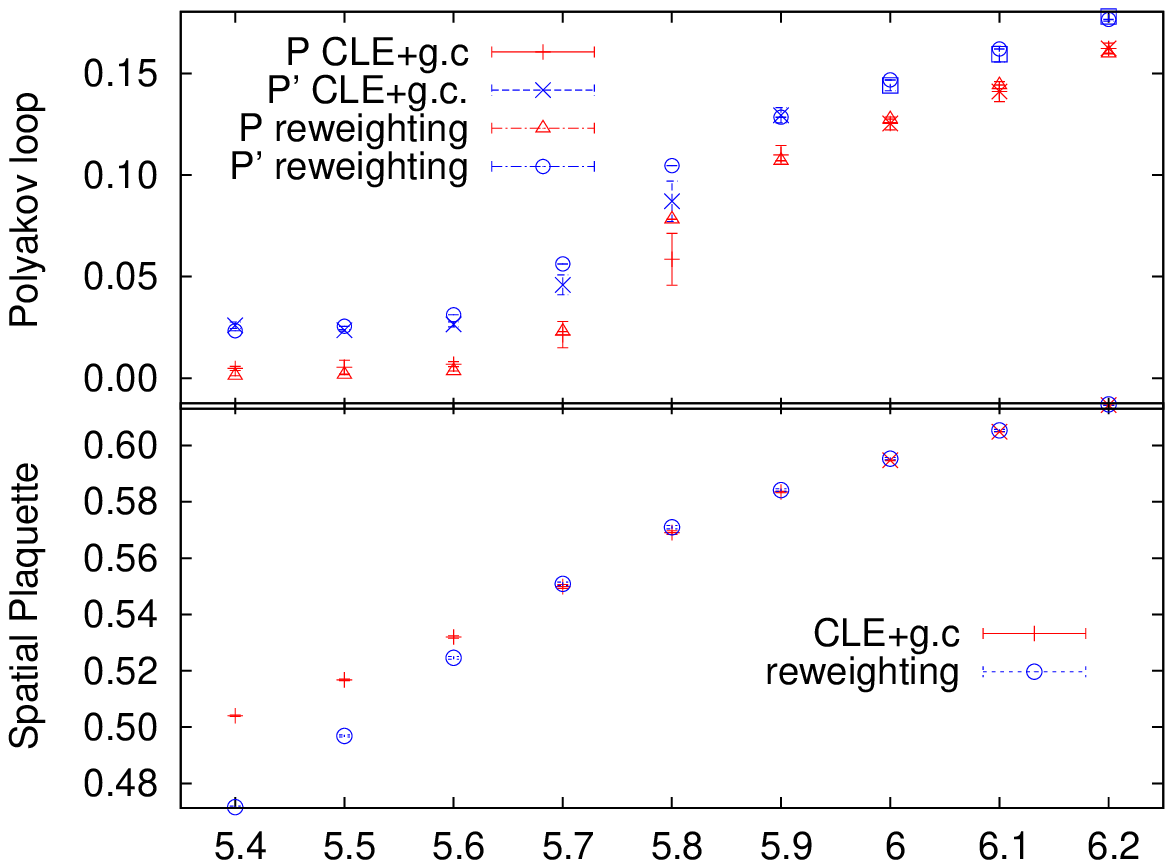}  
\caption{HQCD model: CLE with gauge cooling, plaquette and
Polyakov  loop averages  vs $\mu$ at $\beta=5.9$ (left plot) and vs $\beta$
at $\mu=0.85$ (right plot), $6^4$ lattice, compared with rRW results.  }
\label{hqcd2}
\end{center}
\end{figure}

\begin{figure}[h]
\begin{center}
\includegraphics[width=0.45\columnwidth]{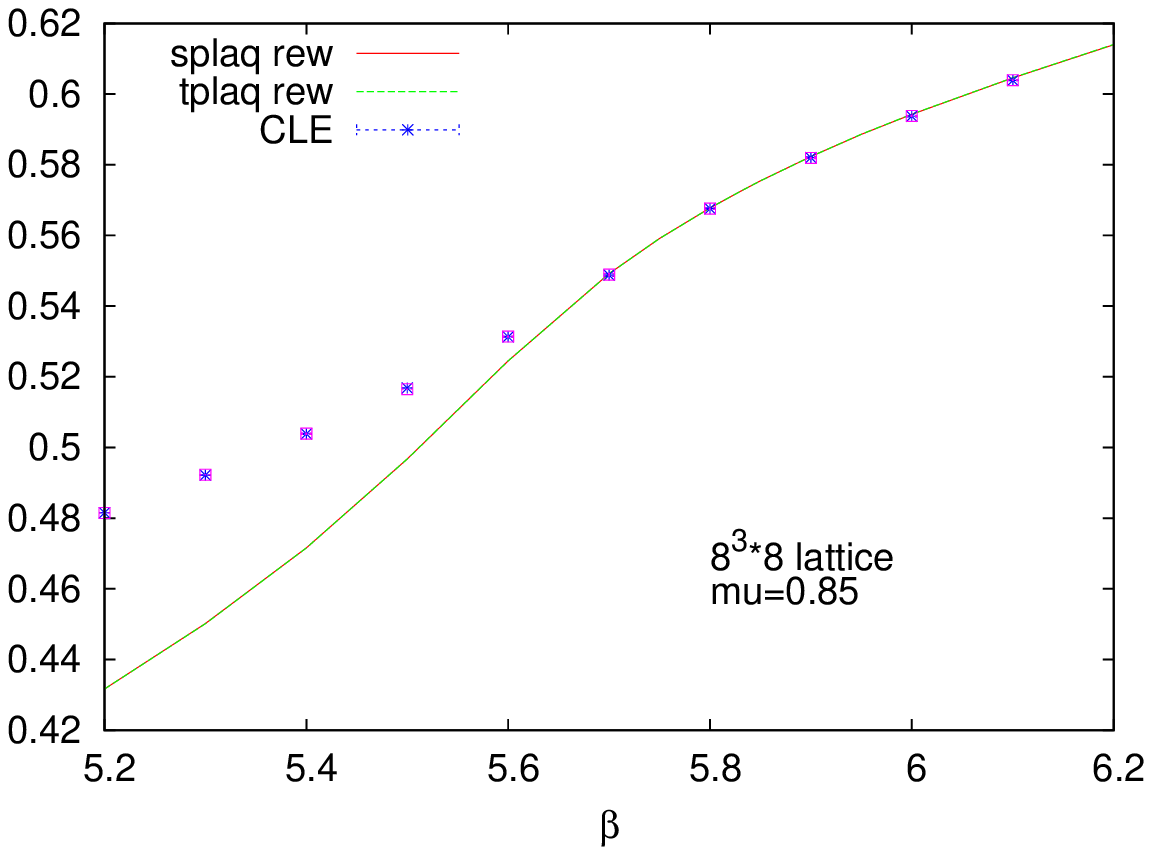}  
\includegraphics[width=0.45\columnwidth]{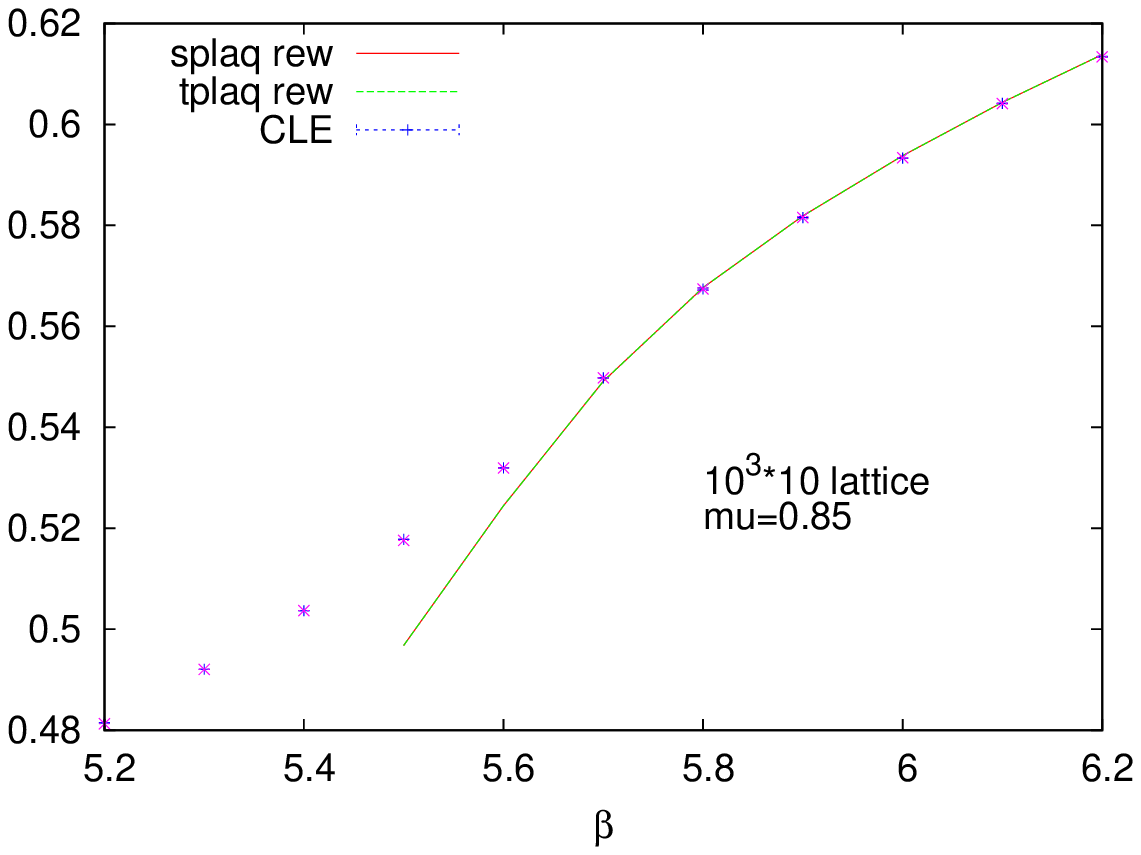}  
\caption{HQCD model: CLE with gauge cooling, plaquette  averages  at 
$\mu=0.85$ compared with the rRW results for two lattice 
sizes: $8^4$ (left) and $10^4$ (right), compare also with previous figure.}
\label{hqcd3}
\end{center}
\end{figure}

In the following we compare the CLE results with those 
from the 
symmetrized version of
rRW - Fig. \ref{hqcd2}.
Both plaquettes and Polyakov loops agree extremely well 
for all values of $\mu$ in the deconfined region 
($\beta =5.9,\ 6^4$ lattice - the large errors affect rRW at large $\mu$). 
At fixed 
$\mu = 0.85,\  6^4$ the agreement persists except  for $\beta < 5.7$,
indicating possible difficulties
 of the CLE. This effect seems, however, to be $\beta$ and not scale 
 dependent, for large lattices we can reach deeply into the confining 
 region (compare the $10^4$ lattice, where the transition is expected 
 at  $\beta \simeq 5.9$) - Fig. \ref{hqcd3}. The excellent agreement 
 between these two completely different methods is a a strong argument 
 for the validity of  both of them in most regions of physial interest.
 For a general review see \cite{review}.


\end{document}